\documentclass[conference]{IEEEtran}
\IEEEoverridecommandlockouts

\usepackage{cite}
\usepackage{amsmath,amssymb,amsfonts}
\usepackage{graphicx}
\usepackage{textcomp}
\usepackage{xcolor}
\usepackage{booktabs}
\usepackage{array}
\usepackage{tabularx}
\usepackage{url}
\usepackage{hyperref} 

\newcommand{\dataset}{LLMail-Inject}
\newcommand{\phaseone}{Phase 1}
\newcommand{\phasetwo}{Phase 2}
\newcommand{\fpr}{FPR}

\def\BibTeX{{\rm B\kern-.05em{\sc i\kern-.025em b}\kern-.08em
    T\kern-.1667em\lower.7ex\hbox{E}\kern-.125emX}}

\begin{document}

\title{Prompt Injection Detection for Email Agents Through Attack Chain Modeling\\
\thanks{A.H. was supported by the Honors Undergraduate Research Fellowship at Eastern Michigan University.}
}

\author{\IEEEauthorblockN{Ahmad Hashmi}
\IEEEauthorblockA{\textit{Department of Computer Science} \\
\textit{Eastern Michigan University}\\
Ypsilanti, MI, USA \\
ahashmi2@emich.edu}
\and
\IEEEauthorblockN{Dhyey Patel}
\IEEEauthorblockA{\textit{Department of Computer Science} \\
\textit{Eastern Michigan University}\\
Ypsilanti, MI, USA \\
dpatel48@emich.edu}
\and
\IEEEauthorblockN{Yunting Yin}
\IEEEauthorblockA{\textit{Department of Computer Science} \\
\textit{Eastern Michigan University}\\
Ypsilanti, MI, USA \\
yyin@emich.edu}
}

\maketitle

\begin{abstract}
Large language model email assistants are particularly vulnerable to indirect prompt injection because untrusted email content can be retrieved into the model context and influence subsequent tool use. Existing prompt injection detectors mainly formulate this problem as binary malicious text classification, which overlooks the important factor that harmful agent behavior often arises through a sequence of stages. We propose a detection framework that models this attack chain by combining a text detector, verifiers specific to each stage, explicit rule-based risk signals, user intent and action consistency analysis, and a logistic decision policy. To support this framework, we derive attack chain labels from prompt injection datasets, evaluate the proposed framework under random splits, temporal phase transfer, conditional stage transfer, cross-dataset transfer, and conduct ablation studies on multiple benchmarks. Results show that random train test splits substantially overestimate robustness under distribution shift, while later tool argument stages are more predictable than earlier stages in the framework. We also show that training on harmless emails that resemble attacks helps reduce false alarms while preserving the ability to detect real attacks. Across five binary benchmarks, our framework achieves a mean
F1 score of 0.406 under the strict threshold setting policy, compared with 0.216 for the strongest of five pretrained detectors evaluated without additional training. These results highlight the value of combining attack stage predictions with checks for conflicts between the user’s request and instructions in retrieved emails. Our experiments also demonstrate the importance of training with challenging benign examples to balance attack detection and false alarms.
\end{abstract}

\begin{IEEEkeywords}
prompt injection, email agents, large language models, AI safety
\end{IEEEkeywords}

\section{Introduction}
Large language model assistants are increasingly integrated with external tools such as email search and calendar management, which enables them to automate complex user workflows\cite{wolflein-etal-2025-llm}. While these capabilities improve productivity\cite{Kobiella_productivity}, they also introduce new security risks. In particular, email assistants routinely retrieve untrusted message content that is incorporated into the LLM context before subsequent reasoning and tool use. Malicious instructions embedded within retrieved emails can therefore influence the model's behavior, causing it to ignore the user's original intent and perform unauthorized actions such as sending emails or leaking sensitive information. This indirect prompt injection threat has emerged as one of the primary security challenges for tool-using LLM agents.

Existing prompt injection research has largely formulated this problem as binary malicious text classification, where the objective is to determine whether a piece of text contains an attack. Although effective for benchmarking detectors, this formulation does not accurately reflect how harmful actions occur in deployed email agents. An attack only succeeds if a sequence of events takes place. In this sequence, malicious content must first be retrieved, evade existing defenses, induce a tool invocation, and ultimately provide unsafe tool arguments that cause an unintended action. Treating this process as a single binary prediction obscures which stages are difficult to detect, which components fail under distribution shift, and where practical defenses should intervene. Furthermore, evaluation based solely on random train test splits may substantially overestimate robustness because future attacks often differ from those observed during training.

Motivated by these observations, we propose a framework for prompt injection detection in email agents. Our proposed framework combines a general text detector, stage verifiers, explicit risk signals based on predefined rules, and user intent and action consistency analysis within a learned logistic meta-classifier. To support this formulation, we derive attack chain labels from the LLMail-Inject benchmark \cite{llmailhf} that represent retrieval, defense bypass, tool invocation, tool-argument correctness, and end-to-end attack success. These labels enable evaluation at multiple points along the attack chain instead of only the final outcome. We further evaluate the framework using temporal phase transfer, conditional stage transfer, cross dataset transfer, threshold setting with benign hard negative examples, and component ablation across multiple public prompt injection benchmarks. Our experiments show that adding action and context signals to all stages raises mean F1 from 0.348 to 0.406. Our proposed configurations outperform the five frozen published detectors in matched mean F1.

The remainder of this paper is organized as follows. Section \ref{sec:review} reviews prior work on indirect prompt injection benchmarks, prompt injection detection models, and defenses for tool-using agents. Section \ref{sec:formulation} formulates the email agent threat model and defines the attack stages considered in this work. Section \ref{sec:data} describes the datasets and attack-chain labels used for training and evaluation. Section \ref{sec:framework} presents the proposed framework together with the experimental methodology. Section \ref{sec:experiments} reports the experimental results, followed by qualitative error analysis in Section \ref{sec:analysis} and discussion of the findings in Section \ref{sec:discussion}. Finally, Section \ref{sec:conclusion} concludes the paper and outlines directions for future work. The implementation and experimental code are publicly available at \href{https://github.com/AcaiLab/Email}{https://github.com/AcaiLab/Email} to support future research.

\section{Related Work}
\label{sec:review}

\subsection{Indirect Prompt Injection Benchmarks for LLM Agents}

Prompt injection is not limited to direct user prompts, but also arises when
LLMs process untrusted external content or tool outputs. Several benchmarks are
proposed to address this problem. BIPIA \cite{bipia} introduces a benchmark for
evaluating prompt injection attacks across email QA, web QA, table QA,
summarization, and code QA, showing that many LLMs are vulnerable to malicious
instructions embedded in external content. This work proposes
boundary-awareness and explicit reminder defenses to reduce attack success.
Tensor Trust \cite{tensortrust} proposes a large benchmark dataset of over
563,000 human-generated prompt injection attacks and 118,000 defenses collected
through an online game, evaluating prompt extraction and prompt hijacking
vulnerabilities while revealing common attack strategies that generalize across
LLMs and some real-world applications. NVIDIA Agentic IPI \cite{nvidiaipi}
provides a benchmark of over a thousand synthetic agentic indirect prompt
injection scenarios across nine domains to evaluate whether tool-using LLM
agents resist malicious instructions embedded in tool outputs that attempt
unauthorized actions. These benchmarks primarily evaluate attack success as a
binary classification problem.

\subsection{Prompt Injection Detection Models}

Research on LLM-based security detection has emphasized the importance of model architecture and evaluation methodology, while identifying dataset scalability and model interpretability as continuing challenges \cite{sheng2025llms, 11012014}. Prompt injection detection models are developed with the goal to identify
malicious inputs before they are processed by an LLM. Ji et al.
\cite{DMPI-PMHFE} propose a prompt injection detection model that combines
semantic embeddings from a pretrained DeBERTa-v3-base model and manually
designed heuristic features that capture known attack keywords and structural
patterns. The two feature channels are concatenated and passed through a neural
classifier to predict whether an input is a prompt injection. Experiments on
multiple benchmark datasets and deployment with several LLMs show higher
detection accuracy and lower prompt injection attack success rates than several
baseline detector models. InjecGuard \cite{notinject} studies over defense,
which is a common failure mode of prompt injection detectors. Over defense
happens when benign inputs containing words like ``ignore'' or ``cancel'' are
falsely flagged as attacks. The paper contributes NotInject, a benign benchmark
with 339 samples designed to test this false positive problem. It also proposes
a detection model, InjecGuard, a DeBERTaV3-base prompt injection classifier
that generates benign examples around biased trigger tokens so the model learns
intent rather than simply reacting to suspicious words. This line of work
motivates our hard-negative evaluation protocol.

\subsection{Defenses for Tool-Using Agents}

As LLMs increasingly interact with external tools, researchers have developed
benchmarks and defenses to secure tool-using agents. AgentDojo \cite{agentdojo}
introduces a benchmark for evaluating prompt injection attacks and defenses in
tool-using LLM agents. It includes 97 agent tasks and 629 security test cases
in email, banking, and travel booking domains. The benchmark shows that current
LLM agents remain vulnerable to prompt injection and highlights the need for
more robust agent architectures and defense mechanisms. Zhan et al. propose
InjecAgent \cite{injecagent}, a benchmark covering attacks that cause direct
user harm or steal private data. The paper evaluates 30 LLM agents and finds
that many tool-using agents are vulnerable, with ReAct-prompted GPT-4 attacked
successfully 24\% of the time, increasing to 47\% when attacker instructions
include an additional hacking prompt. Structural defenses such as StruQ and
CaMeL argue that robust systems require instruction and data separation, as
well as control flow tracking \cite{struq,camel}. These works inspire the
user-intent consistency and action risk assessments in our framework.

\section{Problem Formulation}
\label{sec:formulation}

We model an email assistant as a tool-using LLM that receives a user
request and retrieves email content from it. Our security goal is to prevent untrusted content from causing an unsafe email action. This leads to several design requirements for  our detector: it should distinguish stages of the attack chain, avoid blocking benign text that simply discusses instructions and security, and it should generalize across different benchmarks.

\subsection{Attack Stages}

Prompt injection datasets often provide a binary label: malicious or benign. This is useful for binary classifiers, but it misses whether an email assistant actually retrieved the attack, bypassed a defense, called a tool, or used the target arguments from attackers. These distinctions matter because a message can contain explicit injection language but fail to cause a tool call; another can look like ordinary scheduling text but succeed because it is embedded at the right conversation boundary.

\subsection{Distribution Shift}

Random train and test splits tend to preserve scenario templates and distributions. A detector can therefore learn surface regularities without learning robust attack risk. The \dataset{} data \cite{llmailhf} have two phases, making
it possible to ask whether thresholds and rankings learned from \phaseone{}
generalize to \phasetwo{}. This is closer to how a detector would be deployed in real-world applications.

\subsection{Benign Hard Negatives}

The NotInject benchmark was created around a crucial observation that benign text can
contain trigger words that look like prompt injection \cite{notinject}. If prompt injection
detectors are evaluated only on attacks and ordinary benign text, they may
appear strong while overblocking realistic benign prompts. We test whether this issue appears when combining email attacks with BIPIA indirect attacks \cite{bipia}, and whether explicit hard-negative training can reduce it.

\section{Data and Labels}
\label{sec:data}

\subsection{\dataset{}}

We use the public \dataset{} challenge data released by Microsoft on Hugging
Face \cite{llmailhf}. The dataset contains adaptive submissions against an email
assistant with retrieval, defenses, and objective flags. We parse the objective
fields into six labels:

\begin{enumerate}
\item email retrieved,
\item defense bypassed,
\item tool called,
\item destination correct,
\item content correct,
\item full success.
\end{enumerate}

Table~\ref{tab:dataset} summarizes the cleaned dataset. Full success is 0.814\% in \phaseone{} and 0.337\% in \phasetwo{}, which demonstrates that high accuracy is easy to achieve, while good recall and precision metrics are hard.

\begin{table}[t]
\caption{\dataset{} Phase Summary}
\label{tab:dataset}
\centering
\scriptsize
\begin{tabular}{llrrrrr}
\toprule
Phase & Rows & Retrieved & Bypassed & Tool & Full success \\
\midrule
\phaseone{} & 370,724 & 97.74\% & 50.62\% & 11.11\% & 0.814\% \\
\phasetwo{} & 90,916 & 99.88\% & 48.50\% & 6.31\% & 0.337\% \\
\bottomrule
\end{tabular}
\end{table}

\subsection{Evaluation Datasets}

We use BIPIA as an independent indirect prompt-injection dataset\cite{bipia} for evaluation. BIPIA broadens coverage beyond the LLMail challenge and supports cross dataset transfer tests. We use NotInject as a benign hard-negative
dataset\cite{notinject}. NotInject examples contain suspicious terms or
prompt-like phrasing but are intended to be benign, making them useful for
measuring how much overblocking occurs.

We also use NVIDIA Agentic IPI, PromptShield, ShieldLM, and Neuralchemy
\cite{nvidiaipi,promptshield,shieldlm,neuralchemy} for evaluation. NVIDIA Agentic IPI is attack only. It contains 1,272 indirect injection rows across
nine domains and 40 target tools. PromptShield contributes 43,425
binary detector rows. ShieldLM contributes 54,162 rows spanning benign, direct
injection, indirect injection, and jailbreak categories. Neuralchemy contributes 6,274 detector rows across 31 categories.

\begin{table}[t]
\caption{BIPIA, NotInject, NVIDIA Agentic IPI, PromptShield, ShieldLM, and Neuralchemy Data}
\label{tab:newdata}
\centering
\scriptsize
\begin{tabular}{lrrr}
\toprule
Dataset & Rows & Attack & Benign \\
\midrule
BIPIA generated binary & 176,200 & 175,000 & 1,200 \\
NotInject & 339 & 0 & 339 \\
NVIDIA Agentic IPI & 1,272 & 1,272 & 0 \\
PromptShield & 43,425 & 16,441 & 26,984 \\
ShieldLM & 54,162 & 18,965 & 35,197 \\
Neuralchemy & 6,274 & 3,736 & 2,538 \\
\bottomrule
\end{tabular}
\end{table}

\section{Framework and Experimental Design}
\label{sec:framework}

We conduct experiments to evaluate whether a detector can preserve attack recall under distribution shift while maintaining a low false positive rate on benign hard-negative examples. We answer this by first establishing text detector and
stage modeling baselines, then evaluating the proposed framework against those
baselines on the same held-out datasets and thresholds.

\subsection{Text Detectors}
The main detector is a linear logistic classifier trained with stochastic
gradient descent. Text is represented with TF-IDF word n-grams with
\texttt{ngram\_range=(1,2)}, \texttt{min\_df=3}, \texttt{max\_df=0.98}, unicode
accent stripping, sublinear term frequency, and at most 30,000 features. The
classifier uses an L2 penalty, \texttt{alpha=1e-5}, balanced class weights, and
validation-tuned thresholds chosen to maximize F1. Complement Naive Bayes is
included as a sanity baseline.

For some runs we add handcrafted security features, including counts and indicators for instruction override patterns, role and authority phrases, tool action words,
exfiltration terms, urgency terms, length features, and punctuation statistics. These features did not materially change the early standalone detector conclusions, but they
become useful as explicit risk channels in the proposed framework.

\subsection{Proposed Framework}

The proposed framework implements the staged threat model with six components:

\begin{enumerate}
\item a broad word and character TF-IDF attack detector trained on
LLMail\cite{llmailhf}+BIPIA\cite{bipia}+NotInject\cite{notinject};
\item stage verifiers trained on LLMail \cite{llmailhf} for defense bypass, tool call,
destination correctness, and content correctness, with NotInject examples added
as negative stage examples \cite{notinject};
\item a selective stage policy that downweights or excludes the sparse
full-success verifier when it behaves as a noisy detector feature;
\item action and context consistency checks that compare the user request with
untrusted retrieved content, detecting send-email requests, recipient
introduction, target-address mentions, boundary-control markers, and cases
where untrusted content requests a tool action not requested by the user;
\item rules for instruction override, tool action language and obfuscation;
\item a logistic stacking meta-policy trained on held out validation scores.
\end{enumerate}

\begin{figure}[t]
\centering
\includegraphics[width=\columnwidth,height=0.45\textheight,keepaspectratio]{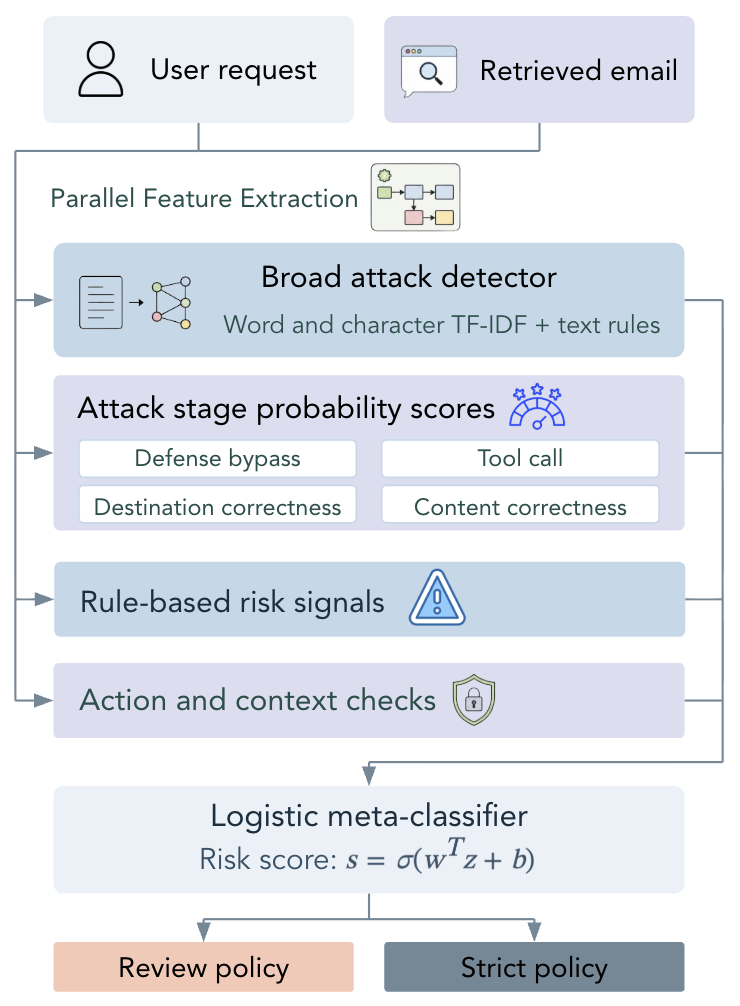}
\caption{Proposed framework workflow. The detector first converts email agent
inputs into broad text, stage verifier, rule, and action and context signals, then
uses a meta policy to select a review or strict decision.}
\label{fig:framework-workflow}
\end{figure}

The framework outputs a single risk score for each input. We evaluate two decision thresholds corresponding to different deployment scenarios. The first threshold is selected to maximize validation F1 and is intended for review-oriented screening. The second threshold is chosen using the NotInject validation set to enforce a low false positive rate and is intended for either block or confirm decisions before potentially unsafe email tool actions are executed. Component ablation determines the final framework configuration for each operating point. For review-oriented screening, the best configuration uses selective stage verification without the full success verifier. Under the strict policy, the configuration combining all stage verifiers with action and context consistency features achieves the highest observed mean F1.

To test whether the stage design actually contributes beyond adding more
features, we run a component ablation. The ablation compares a text only
meta policy, all raw stage verifier scores, selective stage variants, only action and context features, and combined selective stage with action and context features. This protocol shows if stage verifiers or action checks do not add signal, or if a particular stage is noisy.

For comparison with published detectors, we run five text classification models: ProtectAI DeBERTa v1, ProtectAI DeBERTa v2, deepset DeBERTa injection, TestSavant tiny, and Arkaean DistilBERT \cite{protectaiv1,protectaiv2,deepsetinjection,testsavant,arkaean}. The comparison uses the same evaluation datasets \cite{llmailhf,bipia,notinject,nvidiaipi,promptshield,shieldlm,neuralchemy} as the proposed framework.

We also evaluate a zero-shot OpenAI LLM judge gpt-4.1-mini on the full test sets. The judge is prompted to return JSON with an attack probability and a short reason for whether untrusted content attempts prompt injection, instruction override, or unauthorized email-tool control. We evaluate both the default 0.50 threshold and low-\fpr{} thresholds derived from NotInject. When the LLM gives multiple inputs the same probability score, we raise the threshold just above that score if needed to keep false positives within the the false positive limit used to set the threshold.

\subsection{Evaluation Protocols and Metrics}

We evaluate the following protocols:

\begin{itemize}
\item \textbf{Random split}: a sanity check with stratified train, validation, and test
splits.
\item \textbf{Phase transfer}: train on \phaseone{}, and test on \phasetwo{}.
\item \textbf{Conditional stage modeling}: evaluate transitions only among examples where the prerequisite stage occurred.
\item \textbf{Cross-dataset transfer}: train on one or more prompt injection benchmarks and evaluate generalization on held-out benchmarks.
\item \textbf{Proposed framework}: compare the proposed stage and action based framework with a baseline that uses only word- and character-level text features together with rule-based features, under both the validation-F1 threshold and the stricter NotInject threshold.
\item \textbf{Component ablation}: compare text and rule-only, all stage,
selective stage, action and context only, and combined framework variants to
identify which verifier and consistency check signals are helpful.
\item \textbf{Existing detector baselines}: evaluate accessible Hugging Face
prompt-injection detectors.
\item \textbf{LLM judge baseline}: evaluate a zero-shot OpenAI judge gpt-4.1-mini on the full test sets and compare it with the framework.
\end{itemize}

We report F1, precision, recall, ROC-AUC, and false-positive rate.

\section{Results}
\label{sec:experiments}

\subsection{Attack Chain Stage Analysis}

\begin{figure}[t]
\centering
\includegraphics[width=\columnwidth]{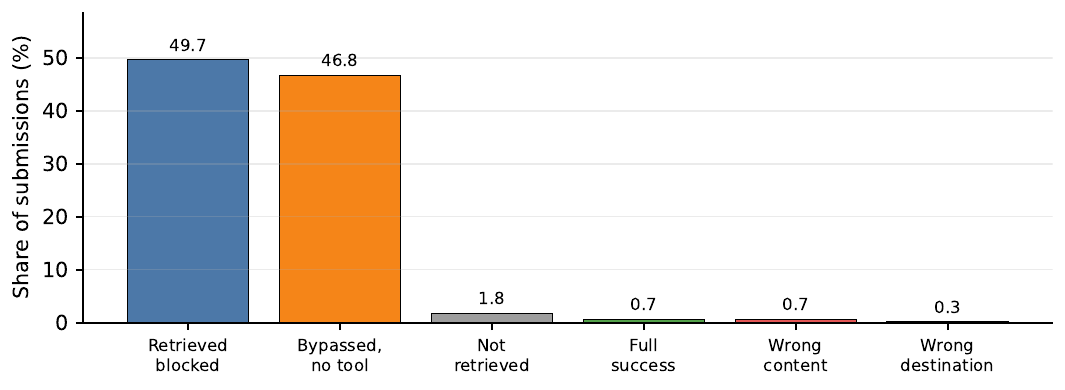}
\caption{Outcomes across the full \dataset{} attack chain. Almost half of
submissions are detected after retrieval, and full success is below 1\% of all
submissions.}
\label{fig:attack-chain}
\end{figure}

Figure~\ref{fig:attack-chain} shows the attack chain distribution. The largest
categories are retrieved-but-detected submissions (49.66\%) and bypassed
submissions with no tool call (46.80\%). Only 0.72\% of all rows are full
successes. Among failures after a tool call, wrong content (0.71\%) is more
common than wrong destination (0.27\%). Table~\ref{tab:generalization} compares random split performance to phase transfer. Random full-success F1 is 0.167 with ROC-AUC 0.825. Under \phaseone{} to \phasetwo{} threshold transfer, full-success F1 falls to 0.000. Even with an oracle \phasetwo{} threshold, full-success F1 is only 0.024. Tool call prediction shows the same pattern: random F1 is 0.527, but phase transfer F1 is 0.077. Defense bypass is more stable but still degrades.

\begin{table}[t]
\caption{Random Split Versus \phaseone{} to \phasetwo{} Transfer}
\label{tab:generalization}
\centering
\scriptsize
\begin{tabular}{lrrrr}
\toprule
Target & Random F1 & Random AUC & Phase F1 & Oracle F1 \\
\midrule
Full success & 0.167 & 0.825 & 0.000 & 0.024 \\
Defense bypass & 0.821 & 0.906 & 0.457 & 0.654 \\
Tool called & 0.527 & 0.892 & 0.077 & 0.137 \\
\bottomrule
\end{tabular}
\end{table}

Our finding shows that the detector learns some regularities in the data, but those regularities do not reliably survive phase shift. The second stage analysis result asks where the chain is actually learnable. Treating every email that looks like an attack as a single binary event overlooks the fact that many messages reach an intermediate stage but do not complete the harmful action.

Table~\ref{tab:stage} reports conditional transition models. Once a tool call has occurred, predicting whether the destination or content is correct is much more stable than predicting whether bypassed text will trigger a tool call. Tool-call-to-destination F1 remains 0.939 under phase transfer, and tool-call-to-content F1 remains 0.814. In contrast, bypass-to-tool-call F1 falls from 0.394 on random split to 0.053 under phase transfer.

\begin{table}[t]
\caption{Conditional Stage-Transition Results}
\label{tab:stage}
\centering
\scriptsize
\begin{tabular}{lrrr}
\toprule
Transition & Random F1 & Phase F1 & Phase AUC \\
\midrule
Retrieved $\rightarrow$ bypassed & 0.814 & 0.413 & 0.455 \\
Bypassed $\rightarrow$ tool call & 0.394 & 0.053 & 0.704 \\
Tool call $\rightarrow$ destination & 0.969 & 0.939 & 0.625 \\
Tool call $\rightarrow$ content & 0.874 & 0.814 & 0.668 \\
Tool call $\rightarrow$ full success & 0.475 & 0.122 & 0.593 \\
\bottomrule
\end{tabular}
\end{table}

This result suggests that email injection defense methods should separate
early stage intent detection from late stage tool argument validation. The latter appears more learnable from text once the assistant has already called a tool.

\subsection{Reducing False Alarms with Benign Training Examples}

Table~\ref{tab:cross} reports cross-dataset results. A detector trained only on
LLMail binary attack or benign proxies \cite{llmailhf} transfers poorly to BIPIA
\cite{bipia} (F1 0.453). This LLMail binary proxy is useful for checking whether detectors recognize LLMail attack text, but its benign side contains only 203 false positive test emails. BIPIA in-domain detection is stronger (F1 0.738, ROC-AUC 0.904), and combining LLMail with BIPIA improves BIPIA test F1 to 0.770.

\begin{table}[t]
\caption{Cross-Dataset and Hard Negative Results}
\label{tab:cross}
\centering
\scriptsize
\begin{tabular}{llrr}
\toprule
Train & Test & F1 / FPR & AUC \\
\midrule
LLMail & BIPIA & 0.453 & 0.572 \\
BIPIA & BIPIA & 0.738 & 0.904 \\
BIPIA & LLMail & 0.747 & 0.780 \\
LLMail+BIPIA & BIPIA & 0.770 & 0.899 \\
LLMail+BIPIA & NotInject & 65.2\% FPR & -- \\
\textbf{LLMail+BIPIA+NotInject} & \textbf{NotInject} & \textbf{0.6\% FPR} & \textbf{--} \\
\textbf{LLMail+BIPIA+NotInject} & \textbf{BIPIA} & \textbf{0.730} & \textbf{0.904} \\
\textbf{LLMail+BIPIA+NotInject} & \textbf{LLMail} & \textbf{0.995} & \textbf{1.000} \\
\bottomrule
\end{tabular}
\end{table}

Adjusting the decision threshold after training to reduce the NotInject \fpr{} to approximately 5\% results in a substantial loss of attack recall. LLMail detector retains only 36.9\% recall on BIPIA attacks, which indicates that post-training threshold adjustment is insufficient to control overblocking in BIPIA-style detectors. Instead, the score distributions of attack examples and benign hard negatives overlap substantially, making it difficult to simultaneously achieve low false-positive rates and high attack recall.

We also compare training with and without NotInject using the first rotation of the data split. When NotInject examples are included in training the text detectors, stage classifiers, and meta policy, the strict selective stage framework with action and context features achieves a mean F1 of 0.451 and incorrectly flags none of the 113 held-out NotInject examples. Without NotInject in training, mean F1 drops to 0.038, and 6 of the 113 benign examples are incorrectly flagged (5.31\%). These results show that targeted benign hard negatives help maintain attack detection while limiting false positives in this setting.

\subsection{Component Ablation}

Table~\ref{tab:component-corrected} shows our main framework experiment. It asks which
parts of the proposed design should remain in the final detector. The table runs a component selection ablation on the same evaluation data using the full available test splits. The ablation compares text and rule scoring, the all stage framework, selective stage verification without the sparse full success verifier, action and context consistency checks without stage scores, and combined stage and action variants.

\begin{table}[t]
\caption{{Framework Component Ablation}}
\label{tab:component-corrected}
\centering
\centering\scriptsize
\resizebox{\linewidth}{!}{%
\begin{tabular}{llrrrr}
\toprule
Variant & Policy & Mean F1 & NotInject \fpr{} & NVIDIA Rec. & BIPIA F1 \\
\midrule
Selective stages & review & $0.643\pm0.002$ & 27.73\% & 100.0\% & 0.745 \\
All stages & review & $0.643\pm0.003$ & 27.14\% & 100.0\% & 0.744 \\
Text/rules only & strict & $0.281\pm0.149$ & 0.88\% & 4.01\% & 0.182 \\
All stages & strict & $0.348\pm0.096$ & 1.77\% & 12.40\% & 0.183 \\
Action/context only & strict & $0.386\pm0.024$ & 0.88\% & 2.33\% & 0.735 \\
Selective stages + action/context & strict & $0.400\pm0.057$ & 2.36\% & 2.33\% & 0.733 \\
\textbf{All stages + action/context} & \textbf{strict} & $\mathbf{0.406\pm0.057}$ & 2.06\% & 2.33\% & 0.732 \\
Nonlinear stacker & strict & $0.347 \pm 0.034$ & 1.47\% & 21.72\% & 0.785 \\
\bottomrule
\end{tabular}}
\end{table}

Table~\ref{tab:component-corrected} reports means and standard deviations over the three NotInject rotations. The strict all-stage action and context variant has the highest mean F1 (0.406), narrowly above selective stages plus action and context (0.400); the difference is small relative to variability. In a same-run comparison, BIPIA F1 rises from 0.183 for all stages without action and context to 0.733 for selective stages plus action and context. The selective action and context policy nevertheless produces 8/339 NotInject false positives (2.36\%, exact 95\% CI: 1.02--4.60\%) and 2.33\% NVIDIA recall. On the fixed first rotation, its five-seed strict mean F1 is $0.411\pm0.067$. A histogram gradient boosted stacker reaches only 0.347 mean F1 overall, despite higher BIPIA F1 and NVIDIA recall, so the linear policy is retained.

\subsection{Comparison with Existing Detectors}

\begin{table}[t]
\caption{Matched Strict Transfer Comparison}
\label{tab:published-corrected}
\centering
\centering\scriptsize
\resizebox{\linewidth}{!}{%
\begin{tabular}{lrrrr}
\toprule
Method & Mean F1 & NotInject FP / 339 & NVIDIA Rec. & BIPIA F1 \\
\midrule
ProtectAI v2 & 0.216 & 13 (3.83\%) & 0.00\% & 0.000 \\
TestSavant tiny & 0.130 & 8 (2.36\%) & 0.00\% & 0.000 \\
ProtectAI v1 & 0.114 & 3 (0.88\%) & 0.00\% & 0.000 \\
Arkaean DistilBERT & 0.094 & 5 (1.47\%) & 0.00\% & 0.000 \\
deepset DeBERTa & 0.014 & 6 (1.77\%) & 0.00\% & 0.013 \\
\textbf{Proposed, all-stage action/context} & \textbf{0.406} & 7 (2.06\%) & \textbf{2.33\%} & \textbf{0.732} \\
\bottomrule
\end{tabular}}
\end{table}

Table~\ref{tab:published-corrected} uses the same test examples and three NotInject data splits that take turns being used for training, threshold setting, and testing. The proposed framework has higher matched strict mean F1 and BIPIA F1 than the five frozen detectors. However, some baselines have lower false positive rates or stronger results on individual datasets, and the proposed framework's 2.33\% NVIDIA recall is low in absolute terms.

\subsection{Comparison with a Zero-Shot OpenAI LLM Judge}

\begin{table}[t]
\caption{Matched Comparison with a Zero-Shot OpenAI LLM Judge}
\label{tab:llmjudge-full}
\centering
\scriptsize
\resizebox{\columnwidth}{!}{%
\begin{tabular}{lrrrr}
\toprule
Method & NotInject \fpr{} & NVIDIA Rec. & BIPIA F1 & LLMail F1 \\
\midrule
Framework, review$^{a}$ & 27.73\% & 100.00\% & 0.745 & 0.665 \\
Framework, strict$^{b}$ & 2.06\% & 2.33\% & 0.732 & 0.665 \\
OpenAI judge, default & 1.47\% & 75.16\% & 0.630 & 0.919 \\
OpenAI judge, strict & 1.77\% & 75.16\% & 0.625 & 0.891 \\
OpenAI judge, ultra-strict & 1.18\% & 73.45\% & 0.565 & 0.810 \\
\bottomrule
\end{tabular}%
}
\vspace{2pt}
\begin{minipage}{\columnwidth}
\scriptsize
$^{a}$Selective stages without action and context, using the
validation-F1 threshold.
$^{b}$All stages plus action and context.
\end{minipage}
\end{table}

Table~\ref{tab:llmjudge-full} compares the revised framework with a zero-shot \texttt{gpt-4.1-mini} judge on the same held-out test rows and the same NotInject data splits for setting detection thresholds and measuring false positives. Under the strict policy, the framework achieves higher BIPIA F1 than the judge (0.732 versus 0.625), with similar observed NotInject false positives (7/339 versus 6/339). However, the judge performs
better on LLMail and NVIDIA and achieves higher mean F1.

The framework nevertheless offers a practical alternative to LLM-based detection. The OpenAI judge requires a hosted LLM inference request for each input, whereas our framework uses sparse TF--IDF features, linear classifiers, explicit rules, and a logistic regression meta-policy. Its detection components can run locally without per input LLM calls, avoiding dependence on an external inference service.

The BIPIA results therefore demonstrate that task-specific attack chain and action and context signals can provide useful detection performance without invoking a generative LLM. This architecture may be attractive when local processing and inference resources are important.

\subsection{Robustness to Padding and Wording Modifications}

We test whether adding benign text can help attacks bypass the selective stage framework with action and context features. Random benign email padding does not reduce recall in our tests, and padding with benign emails with the lowest risk has only a modest effect. However, an attacker who knows the model can select words that strongly lower its risk score. Repeating words with the most negative learned coefficients, including ``please,'' ``creative,'' ``design,'' ``minutes,'' and ``public,'' reduces strict policy recall with 1$\times$ padding placed before or after the attack text. Our results show that carefully chosen benign words can cause the current linear detector to miss attacks.

We also test whether the framework relies on wording specific to LLMail. After removing features that explicitly match the target address or message body, we replace three strings that are specific to the benchmark: the target email address, the word ``confirmation,'' and the tool name. On 500 attacks, recall for the selective stage framework with action and context features decreases slightly, from 0.928 to 0.916. In comparison, recall for the text and rule-only model drops from 0.824 to 0.554. This suggests that checks based on relationships between the user request, email text, and agent output are less sensitive to these particular strings.

\section{Qualitative Error Analysis}
\label{sec:analysis}

We annotated sampled errors from \phasetwo{} full-success prediction and
NotInject false positives. The \phasetwo{} full-success error sample is split
between two dominant patterns. First, 50\% are high-risk near misses, where explicit target artifacts, role markers, or send-email patterns are present, but the
attack chain stops before full success. Second, 44\% are missed conversation boundary attacks, especially meeting confirmation patterns that look like normal scheduling workflows.

NotInject false positives show a different failure mode. The largest categories
are benign technical instructions (30\%), benign security or platform discussion
(26\%), polysemous trigger words (20\%), and benign creative or roleplay prompts
(16\%). In simple terms, the detector often confuses ``text that talks about
instructions or security'' with ``text that is an operational attack.''

\section{Discussion}
\label{sec:discussion}

Our results point to a layered defense strategy. First, early stage detection of
untrusted instructions remains necessary but brittle. Second, tool-call and argument validation are important because later stage labels are more learnable once a tool call occurs. Third, any learned detector should be trained and evaluated with benign hard negatives, because external attack data can produce severe overblocking. The proposed framework is an implementation of this strategy: a broad detector is useful, but it becomes stronger when coupled to selective stage verifiers and more components. The component ablation proves this point. Stage verification is also useful because it separates reliable signals from noisy ones. Action and context consistency checks are most useful when
detection thresholds are set to keep \fpr{} low, where the detector must decide whether untrusted content is trying to cause an action outside the user's request. This means future data collection should prioritize explicit tool action, recipient, and content outcomes. 

\section{Conclusion}
\label{sec:conclusion}
In this work, we view prompt injection detection for email agents as an attack chain modeling problem. By decomposing indirect prompt injection into retrieval, defense bypass, tool invocation, tool argument verification, and end-to-end attack success, we show that different stages exhibit substantially different levels of difficulty and robustness under distribution shift. Our experiments further demonstrate that random train test splits can substantially overestimate detector performance and that benign hard negative examples are essential for evaluating systems in real-world settings.

Motivated by these findings, we propose a stage and action based framework that combines text detection, stage verification, rule-based risk signals, and user intent and action consistency analysis within a learned meta classifier. Across multiple public prompt injection benchmarks, our proposed framework achieves a better balance between attack detection and benign false positives than existing prompt injection detectors. Ablation studies further show that selectively modeling reliable attack stages and incorporating action and context consistency provide the largest improvements, particularly under low false positive requirements.

Our results suggest that securing tool-using LLM email agents requires moving beyond binary prompt classification toward reasoning about how untrusted content propagates through an agent's decision making process. We hope that the proposed attack chain formulation and analyses provide a useful foundation for future research on prompt injection defenses for agentic systems. Future work can potentially evaluate the framework within a live email agent pipeline and expand the diversity and scale of benign hard negative examples.

\bibliographystyle{IEEEtran}
\bibliography{references}

\end{document}